\documentclass[aps,pre,superscriptaddress,unsortedaddress,twocolumn,showpacs]{revtex4}

\usepackage{amsfonts}

\usepackage{graphicx}
\usepackage{epstopdf}
\usepackage{amsmath}
\usepackage{amssymb}
\usepackage{verbatim}
\usepackage[ulem=normalem]{changes}
\usepackage{booktabs}
\usepackage{natbib}

\begin{document}

\title{Realization of nonequilibrium thermodynamic processes using external colored noise}

\author{Pau Mestres}
\affiliation{ICFO $-$ Institut de Ci\`encies Fot\`oniques, Mediterranean Technology Park,  Av. Carl Friedrich Gauss, 3, 08860, Castelldefels (Barcelona), Spain.}
\author{Ignacio A. Martinez}
\affiliation{ICFO $-$ Institut de Ci\`encies Fot\`oniques, Mediterranean Technology Park,  Av. Carl Friedrich Gauss, 3, 08860, Castelldefels (Barcelona), Spain.}
\author{Antonio Ortiz-Ambriz}
\affiliation{ICFO $-$ Institut de Ci\`encies Fot\`oniques, Mediterranean Technology Park,  Av. Carl Friedrich Gauss, 3, 08860, Castelldefels (Barcelona), Spain.}
\affiliation{Photonics and Mathematical Optics Group, Tecnol\'ogico de Monterrey, 64849, M\'exico.}
\author{Raul A. Rica}
\email{rul@ugr.es}
\affiliation{ICFO $-$ Institut de Ci\`encies Fot\`oniques, Mediterranean Technology Park,  Av. Carl Friedrich Gauss, 3, 08860, Castelldefels (Barcelona), Spain.}
\author{Edgar Roldan}
\email{edgar.roldan@fis.ucm.es}
\affiliation{ICFO $-$ Institut de Ci\`encies Fot\`oniques, Mediterranean Technology Park,  Av. Carl Friedrich Gauss, 3, 08860, Castelldefels (Barcelona), Spain.}
\affiliation{GISC $-$ Grupo Interdisciplinar de Sistemas Complejos. Madrid, Spain.}

\medskip

\begin{abstract}
We investigate the dynamics of single microparticles immersed in water that are driven out of equilibrium in the presence of an additional external colored noise. As a case study, we trap a single polystyrene particle in water with optical tweezers and apply an external electric field with flat spectrum but a finite bandwidth of the order of kHz. The intensity of the external noise controls the amplitude of the fluctuations of the position of the particle, and therefore of its effective temperature. Here we show, in two different nonequilibrium experiments, that the fluctuations of the work done on the particle obey Crooks fluctuation theorem at the equilibrium effective temperature, given that the sampling frequency and the noise cutoff frequency are properly chosen. Our experimental setup can be therefore used to improve the design of microscopic motors towards fast and efficient devices, thus extending the frontiers of nano machinery. 

 \end{abstract}

\pacs{05.70.Ln, 05.20.-y, 05.40.-a,42.50.Wk}

\maketitle

\section{Introduction}

The thermodynamics of small systems is strongly affected by the thermal fluctuations of the surroundings~\cite{seifert2012stochastic}. Although often looked as an unwanted source of noise, fluctuations also bring to life phenomena such as stochastic resonances~\cite{volpe2008stochastic}, temporal violations of the Second Law of Thermodynamics~\cite{wang2002experimental} or the possibility to build engines at the micro scale of a greater efficiency than that of their macroscopic counterparts~\cite{blickle2012realization,yasuda1998f}. 

At the microscale, the amplitude of the fluctuations of the thermodynamic quantities depends on the temperature of the environment. Unfortunately, temperature control has remained challenging due to the difficulties found to isolate microscopic systems and to the presence of convection effects in fluids~\cite{mao2005temperature,millen2014nanoscale}. In Ref.~\cite{mao2005temperature}, a thermal collar was attached to an objective in order to heat up a sample fluid. In a different approach, a laser line matching the absorption peak of water was used to heat the sample uniformly, thus avoiding convection~\cite{blickle2012realization}. Although the aforementioned methods were proved capable of controlling the temperature, they were limited to increase the temperature of the sample in a range of tens of Kelvins.

To overcome the short range of accessible temperatures, it has been suggested that random forces can mimic a thermal bath for colloidal particles~\cite{sekimoto-bible,gomez2010steady,martinez2013effective}. In fact, the existence of fluctuations in the small scale is due to neglected degrees of freedom in the description of the system~\cite{sekimoto-bible}. In other words, thermal fluctuations are produced by the constant exchange of energy between the system under consideration and the $\sim N_A$ molecules of the surrounding environment, $N_A$ being the Avogadro number. In Ref.~\cite{gomez2010steady}, it is shown that an interpretation of random forces as a source of work results in a failure of Crooks fluctuation theorem~\cite{crooks1999entropy}. On the other hand, if random forces are considered as a heat source, one finds a behavior equivalent to that of a colloidal particle in a thermal bath with equilibrium temperatures ranging on the thousands of degrees~\cite{martinez2013effective}. Therefore, it is possible to argue that an external source of noise can thought of as a {\em virtual} thermal bath~\cite{sekimoto-bible,martinez2013effective}. The latter interpretation was experimentally tested in~\cite{martinez2013effective}, were  deviations between the effective temperatures in equilibrium and in nonequilibrium processes were found. This lack of correspondence makes dubious whether using random forces is suitable to mimic a thermal bath.

Optical tweezers offer a robust and versatile platform for micromanipulation~\cite{ashkin1970acceleration,evers2013particle,simon2013transport,gieseler2014dynamic,millen2014nanoscale,volpe2014speckle} and for the study of the thermodynamics of systems where fluctuations cannot be neglected~\cite{sekimoto-bible,seifert2012stochastic,roldan2014irreversibility}. Interestingly, it can be easily combined with other experimental techniques in order to broaden its applicability. For example, optical trapping has been combined with Raman spectroscopy~\cite{raj2012mechanochemistry,rao2013direct} or fluorescence microscopy~\cite{gross2011quantifying} to study single molecule biology. When the trapped objects are charged, the application of electric fields can be used to perform single particle electrophoresis~\cite{lu2012single,semenov2013electrophoretic,strubbe2013electrophoretic,jonas2008light} or study the fundamental laws of thermodynamics at small scales~\cite{Roldan2014Universal,roldan2014measuring,martinez2014realization}.

In this article, we extend the use of random forces to mimic a thermal bath for a colloidal particle undergoing nonequilibrium processes in an optical trap. In particular, we analyze the validity of the interpretation of a noisy electric force as a heat bath  in out-of-equilibrium dragging and expansion-compression processes. With data from both experiments and numerical simulations, we demonstrate that the observed mismatch between equilibrium and nonequilibrium kinetic temperatures can be caused by an inappropriate sampling during the experiment. We show that the fluctuations of thermodynamic quantities are very sensitive to the sampling frequency and to the actual properties of the external noise, which in practice will always be colored.

This article is organized as follows. In Sec.~\ref{sec:stoch_energetics} we discuss the theoretical extension of Crooks fluctuation theorem to small systems driven by external random forces. In Sec.~\ref{sec:experiment} we describe the experimental setup and the protocols that we use to implement nonequilibrium processes with external colored noise. In Sec.~\ref{sec:results} we show and discuss the experimental results obtained in the implementation of two different nonequilibrium processes with a trapped colloidal particle, proposing an optimal sampling rate to realize nonequilibrium processes with our setup. Section~\ref{sec:conclusions} shows the concluding remarks of our work.

\section{Crooks fluctuation theorem under random driving}
\label{sec:stoch_energetics}

Let us consider a Brownian colloidal particle that moves in one dimension $x$ and is immersed in a thermal bath at temperature $T$. The particle is trapped with a quadratic potential centered at the position $x=x_0$, $U(x,x_0,\kappa) = \frac{1}{2}\kappa (x-x_0)^2$, $\kappa$ being the stiffness of the trap. The position of the particle obeys the overdamped Langevin equation~\cite{Langevin1908},
\begin{equation}\label{langevin}
\gamma\dot x(t) = -\kappa(t)\left[ x(t)-x_0(t)\right] + \xi(t)+\eta(t).
\end{equation}
where $\gamma$ is the friction coefficient and both the stiffness and the position of the trap can change with time $t$. The term $\xi(t)$ is a stochastic force that accounts for the random impacts of the molecules in the environment with the particle, responsible for its Brownian motion, which is modelled by a Gaussian white noise of zero mean $\langle \xi(t)\rangle =0$ and correlation $\langle \xi(t)\xi(t')\rangle = 2\gamma kT\delta(t-t')$. We also consider the possibility that an additional external random force $\eta(t)$ is applied to the particle, which satisfies $\langle \eta(t)\rangle=0$ and $\langle \eta(t)\eta(t')\rangle=\sigma^2\Gamma(t-t')$, where $\Gamma(t-t')$ is the correlation function of the force and $\sigma$ its amplitude. In the general case, this correlation function is different from a delta function.  

In the absence of external forces and being $\kappa$ and $x_0$ fixed at a constant value, the fluctuations of the position of the particle are Gaussian-distributed. The amplitude of these fluctuations depends on the temperature of the surroundings, as predicted by equipartition theorem, $\kappa \langle (x-x_0)^2\rangle = kT$, where the brackets
denote steady-state averaging. If we also include the external random force $\eta(t)$, equipartition theorem allows us to define a kinetic temperature of the particle from the amplitude of its motion,
\begin{equation} \label{eq: equilibrium}
T_{\rm kin}=\frac{\kappa\langle(x-x_0)^2\rangle}{k}.
\end{equation}

If $\eta(t)$ has the same nature as the Brownian force, i.e., it is described by a Gaussian white noise, then $T_{\rm kin}=T+\dfrac{\sigma^2}{2k\gamma}$. Otherwise, the relation between $T_{\rm kin}$ and the noise intensity is more complex, and depends on the characteristic timescales of the particle~\cite{martinez2013effective}. Anyhow, it always verifies $T_{\rm kin }\ge T$~\cite{martinez2013effective,roldan2014measuring}. 

Let us now consider a thermodynamic process along which an external agent can control the energy of the Brownian particle via a control parameter $\lambda$ that can be arbitrarily switched in time, for instance the stiffness of the trap. The duration of the process is $\tau$ and the control parameter follows a protocol $\{\lambda_t\}_{t=0}^{\tau}\equiv\{\lambda(t)\}_{t=0}^{\tau}$. For convenience, we also consider the {\em time-reversal} process, described by $\{\tilde\lambda_t\}_{t=0}^{\tau} = \{\lambda_{\tau-t}\}_{t=0}^{\tau}$. We assume that the system is initially in canonical equilibrium state and is allowed to relax to equilibrium at the end the process. The position of the particle is random and describes a stochastic {\em trajectory}, $\{x_t\}_{t=0}^{\tau}\equiv\{x(t)\}_{t=0}^{\tau}$. Along the process, the external agent exerts work on the particle, which depends on the trajectory of the particle~\cite{sekimoto-bible},
\begin{equation}\label{work}
W=\int^{\tau}_{0} \dfrac{\partial U(x_t,\lambda_t)}{\partial \lambda_t} \circ d\lambda_t,
\end{equation}
where $\circ$ denotes Stratonovich product~\cite{sekimoto-bible}.

The average of the work over many different realizations yields the classical result $\langle W\rangle \geq \Delta F$, where $\Delta F$ is the free energy difference between the final and initial (equilibrium) states of the system~\cite{sekimoto-bible}. The fluctuations of the work are not symmetric upon time-reversal of the protocol if the system is driven out of equilibrium, as first shown by Crooks~\cite{crooks1999entropy}. For an arbitrarily far-from-equilibrium process, the work distribution of the {\em forward} process $\rho(W)$  is related to the distribution of the work in the {\em backward} (time-reversal) process, $\tilde \rho(W)$,
\begin{equation}
\frac{\rho(W)}{\tilde \rho(-W)}=\exp\left(\frac{W-\Delta F}{kT}\right).
\label{eq:CrooksLasVegas}
\end{equation} 
One can also define the following {\em asymmetry function}~\cite{garnier2005nonequilibrium},
\begin{equation}
\Sigma (W)\equiv\ln \frac{\rho(W)}{\tilde \rho(-W)},
\label{eq:asymmetryfunction}
\end{equation}
which measures the distinguishability between the forward and backward work histograms. Crooks fluctuation theorem (CFT) can be rewritten in terms of the asymmetry function,
\begin{equation}
\Sigma (W) =  \frac{W-\Delta F}{kT}.
\end{equation}

In Ref.~\cite{collin2005verification}, CFT was tested experimentally in DNA pulling experiments. In the presence of an external random force, CFT is not satisfied if the external force is considered to exert work on the particle~\cite{gomez2010steady}.  If, however, the energy transferred by the random force is considered as heat, the following CFT is satisfied
\begin{equation}\
\Sigma (W) =  \frac{W-\Delta F}{kT_c},
\label{eq:CFTcolor}
\end{equation}
where $T_c$ is an effective nonequilibrium temperature called {\em Crooks temperature}. Equation~\eqref{eq:CFTcolor} implies that $T_c$ can be calculated from the slope of $\Sigma(W)$ as a function of $W$. The value of the effective Crooks temperature depends strongly on the properties of the external noise. In general, $T_c$ and its equilibrium counterpart, $T_{\rm kin}$ do not coincide. However, if the external force is an external Gaussian white noise, $T_c = T_{\rm kin}$~\cite{martinez2013effective}. 

%
%
%Finally, a more general expression of the fluctuation theorem involving the entropy writes as follows:
%
% 
%
%\begin{equation}\label{eq: stot}
%\Delta S_{tot}=\Delta S_{sh}-\dfrac{Q}{T} 
%\end{equation}
%
%Where $\Delta S_{sh}$ corresponds to the Shanon entropy of the system: 
%\begin{equation}\label{eq: sshanon}
%\Delta S_{sh}= -k_b\left( log\left( \dfrac{\rho(x(t_f))}{\rho(x(t_o))} \right) \right)
%\end{equation}
%with $\rho(x)$ calculated from the initial and final steady state position distributions of the  process. And the heat $Q$ is computed as:
%
%\begin{equation}\label{eq: heat}
%Q=\int^{x(\tau)}_{x(0)}\dfrac{\partial V(x(t),\lambda(t)))}{\partial x} \circ \text{d}x (t)
%\end{equation}
%

\section{Experimental Methods}
\label{sec:experiment}
\subsection{Experimental setup}
\label{sec:onlyexperiment}

\begin{figure} 
 \includegraphics[width=6.5cm]{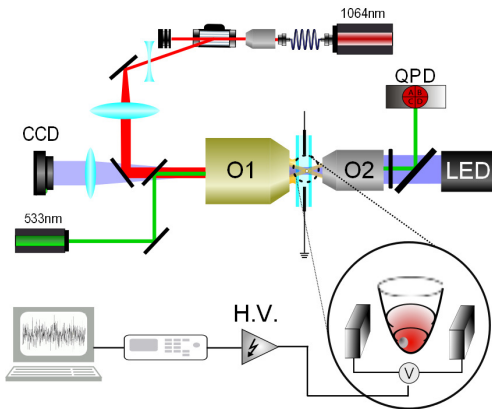}
 \caption{Experimental setup described in Sec.~\ref{sec:onlyexperiment}. A polystyrene sphere of radius $R=500\,\rm nm$  is immersed in a micro-fluidic chamber filled with water and trapped with an infrared laser using a high numerical aperture objective. Random forces are exerted using a Gaussian white noise process applied to two electrodes placed at the two ends of the chamber. The position is detected projecting the forward scattered light of a green laser in a quadrant photodiode. }
 \label{fig: figure1}
\end{figure}

Our experimental setup, shown in Fig.~\ref{fig: figure1}, was previously described in Ref.~\cite{martinez2013effective}. We use a 40x objective to collimate the laser beam from a single-mode fiber laser (ManLight ML10-CW-P-OEM/TKS-OTS, 3W maximum power) and send it through an Acousto Optic Deflector (AOD). After the AOD, the beam is expanded with 2 lenses that also conjugate the center of the AOD crystal with the entrance of a 100x objective (Nikon, CFO PL FL NA1.3) (O1) that creates the field gradient for the optical trap.

For position detection a $\lambda=532\,$nm fiber laser is expanded with a 10x objective and sent through the same objective of the optical trap (O1). The forward scattered light is collected with a 20x (O2) objective and sent to a Quadrant Photodiode (QPD) with 50kHz acquisition bandwidth and nanometer accuracy.

Our sample consists of  polystyrene microspheres of diameter $D=(1.00\pm0.05)\,\mu$m (PPs-1.0, G.Kisker-Products for Biotechnology) injected into a custom made electrophoretic chamber that can be moved using a piezoelectric stage (Piezosystem Jena, Tritor 102), see Ref.~\cite{tonin2010electrophoretic}. 

The intensity and position of the trap center  can be controlled by changing the modulation voltage ($V_\kappa$) and the driving voltage of the AOD ($V_{\rm AOD}$), respectively. In order to know the position of the trap and its stiffness we need to obtain the calibration factors between $V_{\kappa}$ and $\kappa$ as well as between $V_{\rm AOD}$ and $x_0$.  First, we measure $\kappa$ by fitting the Power Spectral Density (PSD) of the position of a trapped bead to a Lorentzian function~\cite{berg2004power} at different values of $V_{\kappa}$. Secondly, the calibration of $x_0$ as a function of $V_{\rm AOD}$ is obtained  from the analysis of the average position of a trapped bead in equilibrium, for different values of $V_{\rm AOD}$ (data not shown).

 The external random electric field is generated from a Gaussian white noise process. The sequence was obtained using independent random variables as described in Ref.~\cite{martinez2013effective}. The signal from the generator is amplified and applied directly to the two electrodes connected at the two ends of the electrophoretic chamber. Notice that the noise spectrum is flat up to a cutoff frequency $f_{\rm co} = 10\,\rm kHz$ (given by the amplifier), which exceeds by one order of magnitude the cutoff frequency used in Ref.~\cite{martinez2013effective}. 

\subsection{Protocols}
\label{sec:protocols}
We consider two different nonequilibrium processes. First, we study the dynamics of a microscopic sphere in an optical trap that is dragged at constant velocity. Next, we realize a process where the trap center is held fixed but the stiffness of the trap is changed with time. In both cases, we perform a quantitative study of the nonequilibrium work fluctuations of the processes and their time reversals using CFT, as discussed in Sec.~\ref{sec:stoch_energetics}.  

\subsubsection{Dragged trap}

\begin{figure}
 \centering
 \includegraphics[width=6cm]{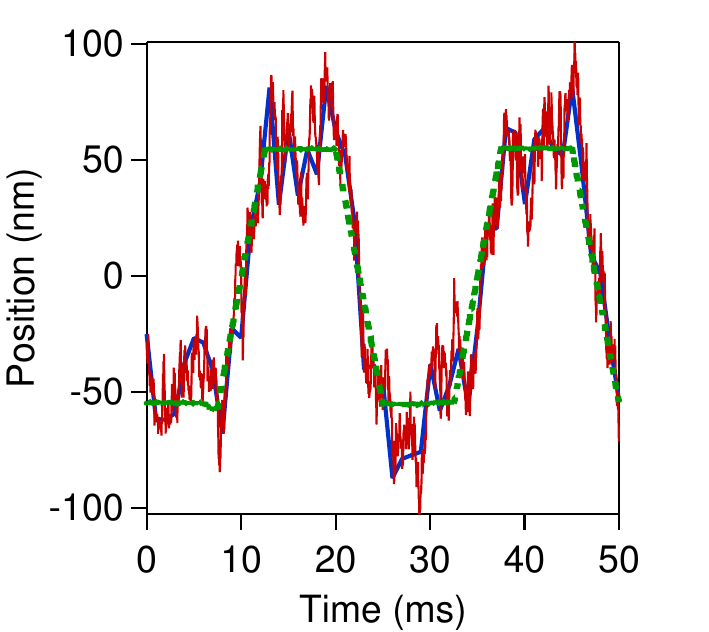}
 \caption{Position of the trap (green dashed curve) as a function of time and time traces of the position of the particle sampled at $1$ kHz (blue curve) and $10$ kHz (red curve) in the dragging experiment.}
 \label{fig: figure2}
\end{figure}

Our first case of study consists of a particle that is driven out of equilibrium by dragging the optical trap of stiffness $\kappa=(18.0\pm 0.2)\,\rm pN/\mu$m at constant speed $v=22\,\rm nm/ms$. The protocol is shown in Fig.~\ref{fig: figure2} together with a time series of the position of the particle sampled at different acquisition frequencies. First, the trap is held fixed with its center at $x_0=-55\,\rm nm$ during $\tau_1 = 7.5\,\rm ms$. Then the trap center is displaced in the $x-$axis at a constant velocity from $x_0=-55\,\rm nm$ to $x_0=55\,\rm nm$ in a time interval of $\tau_2=5\,\rm ms$. The bead is then allowed to relax to equilibrium by keeping the trap center fixed at $x_0=55\,\rm nm$ for $\tau_1 = 7.5\,\rm ms$ before the trap is moved back from $x_0=55\,\rm nm$ to $x_0=-55\,\rm nm$ in $\tau_2=5\,\rm ms$. The duration of each cycle is $\tau=25$ ms and every cycle is repeated $12000$ times, that is, the total experimental time was $300$ s. Every $300\,\rm s-$cycle is repeated for different values of the amplitude of the random force, starting with the case where no external force is applied.\\

%We notice that the time spent by the trap at a fixed position is enough to let the particle relax to equilibrium, since it exceeds one order of magnitude the position relaxation time, $\tau_1 = 7.5\,\rm ms > \tau_r = \gamma / \kappa = 0.5 \rm ms$, $\gamma=8.4$ pN ms/$\mu$m being the friction coefficient. 
The relaxation time of the position of the particle is $\tau_r = \gamma / \kappa = 0.5\,\rm ms$ where $\gamma=8.4$ pN ms/$\mu$m is the friction coefficient. The time spent by the trap in the fixed stage of the protocol then exceeds by one order of magnitude the calculated relaxation time, which should be enough to make sure the particle reaches equilibrium before the next step of the protocol.
In the dragging steps, the viscous dissipation is of the order of $\langle W_{\rm diss}\rangle \sim \gamma vL$, where $L=110\,\rm nm$ is the distance travelled by the trap, which yields $\langle W_{\rm diss}\rangle \sim 20\,\text{pN\,nm} \simeq 5\,kT$ indicating that the work dissipation cannot be neglected and the system is    therefore out of equilibrium.

In every cycle of the protocol, we calculate the work done on the particle in the forward and backward process using Eq.~\eqref{work}. In this case, the control parameter is the position of the trap center, $\lambda = x_0$, and therefore the work is calculated as
 \begin{equation}
 W=\int \frac{\partial U}{\partial x_0} \circ dx_0(t) =\int -\kappa (x(t)-x_0(t))\circ dx_0 (t),
 \end{equation}
  for every realization of the forward and backward processes. 

\subsubsection{Isothermal compression and expansion}

\begin{figure} 
 \includegraphics[width=7cm]{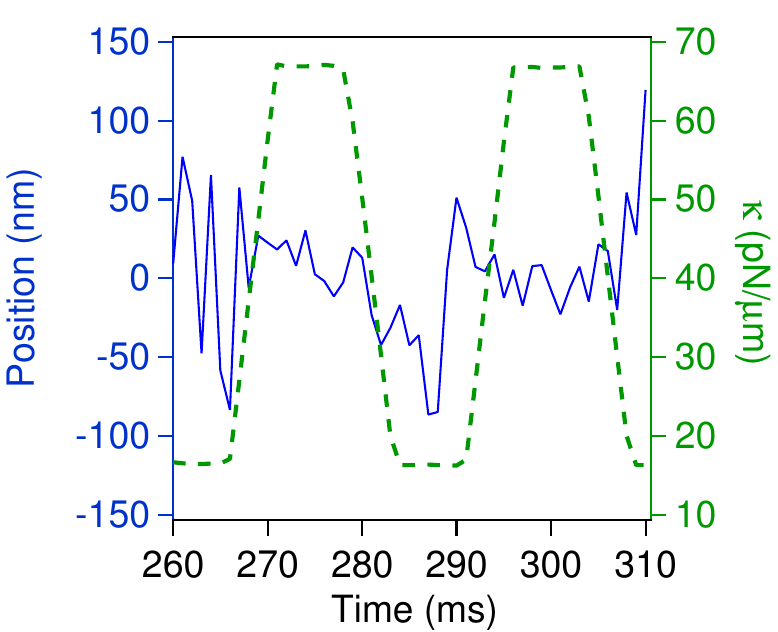}
 % positions.pdf: 0x0 pixel, 300dpi, 0.00x0.00 cm, bb=
 \caption{Position of the particle (blue line, left axis) and trap stiffness (green line, right axis)  as functions of time in the isothermal compression-expansion cycle. Sampling rate, $f=1\,\rm kHz$.}
 \label{fig:breathprotocol}
\end{figure}

As a second application of our technique, we analyze a different thermodynamic process consisting in a ``breathing" harmonic potential, where the trap center is held fixed but its stiffness is changed with time from an initial $\kappa_{\rm ini}$ to a final $\kappa_{\rm fin}$ value. Since the stiffness of the trap can be thought of as the inverse characteristic volume of the system, $\kappa\sim 1/V$~\cite{blickle2012realization}, such process is equivalent to an isothermal compression or expansion. At odds with the dragging process, in this case the free energy changes along the process, yielding $\Delta F = kT_{\rm kin}\ln\sqrt{\kappa_{\rm fin}/\kappa_{\rm ini}}$~\cite{sekimoto-bible,roldan2014measuring}.

In the experimental protocol shown in Fig.~\ref{fig:breathprotocol}, the trap is initially held fixed with stiffness $\kappa_1=(16.5\pm 0.2) \rm pN/\mu$m for $\tau_1=3.5$ ms. Then, the system is isothermally compressed by increasing the stiffness up to $\kappa_{2}=(66.8\pm 0.2)pN/\mu$m in $\tau_2=2.5$ ms. Further, the particle is allowed to relax to equilibrium for $\tau_1=3.5$ ms with the trap stiffness held fixed at $\kappa_{2}$ before the system is isothermally expanded from $\kappa_2$ to $\kappa_1$ in $\tau_2=2.5$ ms. Every cycle lasts $\tau=2(\tau_1 + \tau_2) = 12\,\rm ms$ and is repeated $24000$ times for different values of the amplitude of the external random force.

For every isothermal compression (forward process) and expansion (backward process), we measure the work done on the particle as 
\begin{equation}
W=\int \frac{\partial U}{\partial \kappa} \circ d\kappa(t) = \int \frac{1}{2}x^2(t) \circ d\kappa(t),
\label{eq:workbreath}
\end{equation}
were the control parameter is the trap stiffness in this case [$\lambda=\kappa$ in Eq.~\eqref{work}].

\section{Results and discussion}
\label{sec:results}

%We now study if our setup can be used to implement thermodynamic nonequilibrium processes of different nature.
We now discuss the results obtained when implemented the two different nonequilibrium processes described in Sec.~\ref{sec:protocols}. For both processes, we perform a quantitative study of the nonequilibrium work fluctuations of the processes and their time reversals using CFT, as discussed in Sec.~\ref{sec:stoch_energetics}.

\subsection{Dragged trap}

\begin{figure}
 \includegraphics[width=7cm]{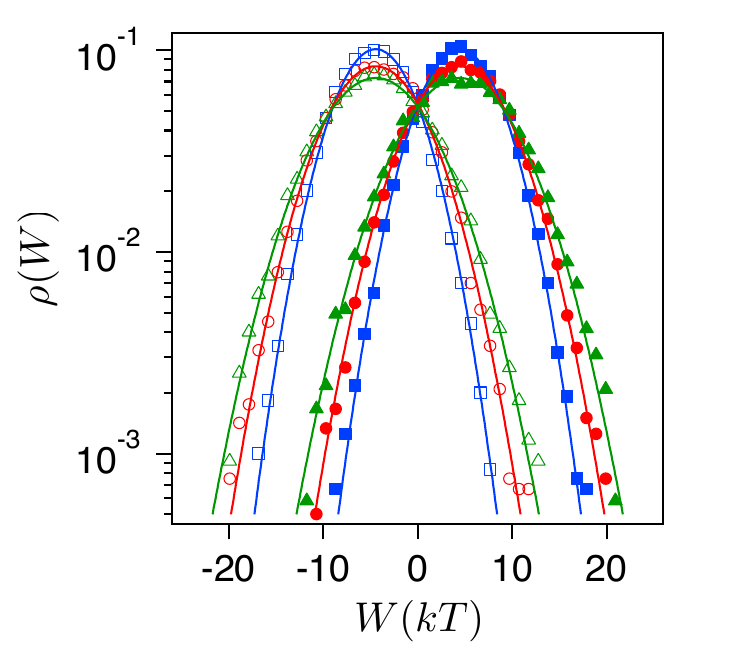}
 \caption{Work distributions in the forward ($\rho(W)$, filled symbols) and backward ($\tilde\rho(-W)$, open symbols) dragging experiments depicted in Fig.~\ref{fig: figure2}. Different symbols and colors correspond to different noise intensities, yielding the following values of the Crooks temperature:  $T_{c}=525\,\rm K$ (blue squares)  $T_{c}=775\, \rm K$ (red circles) and $T_{c}=1010\, \rm K$ (green triangles).  Solid lines are the theoretical values of the work distributions obtained for the same values of kinetic temperatures. Work was calculated from trajectories sampled at $f=10\,\rm kHz$.}
 \label{fig: figure3}
\end{figure}

Figure \ref{fig: figure3} shows the work distributions at different noise intensities for both forward and backward dragging processes. When increasing the noise amplitude, the average work remains constant but the variance  increases. Since in this process, the free energy does not change, $\Delta F=0$, then the average work coincides with the average dissipation rate $\langle W\rangle = \langle W_{\rm diss}\rangle$. Therefore, the addition of the external random force does not introduce an additional source of dissipation and can be treated as a heat source.  The work distributions at different noise amplitudes fit to theoretical Gaussian distributions obtained from Refs.~\cite{saha2009entropy,martinez2013effective} using as the only fitting parameter the nonequilibrium Crooks temperature, that enters in the asymmetry function, as indicated by Eq.~\eqref{eq:CFTcolor}.

% fit well with the theoretically-predicted  To compare these results with the theory, we also plot the analytical work distributions at the same values of $T_C$ as described in \cite{saha2009entropy} and we see a perfect agreement.

\begin{figure}
 \includegraphics[width=7cm]{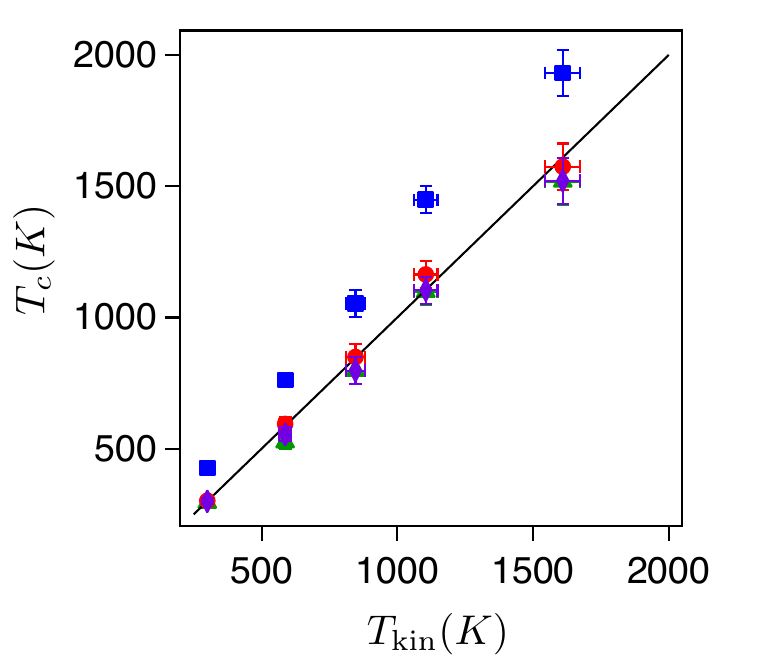}
 \caption{Effective nonequilibrium kinetic temperature, $T_c$,  vs effective equilibrium kinetic temperature, $T_{\rm kin}$, for different amplitudes of the external noise. Different symbols correspond to results obtained for different sampling rates: $1\,$kHz (blue squares), $2\,$kHz (red circles), $5\,$kHz  (green triangles) and $10\,$kHz (magenta diamonds). Solid black line corresponds to $T_{c}=T_{\rm kin}$. Error bars represent statistical errors with a statistical significance of $90\%$.}
 \label{fig: figure4}
\end{figure}

As already discussed, if we want this technique to be applicable to the design of nonequilibrium thermodynamic processes, one would require that the equilibrium and nonequilibrium kinetic temperatures, that is $T_{\rm kin}$ and $T_c$, coincide within experimental errors. However, some discrepancies were found in Ref.~\cite{martinez2013effective} when the sampling frequency was changed, and their origin could not be fully understood. 
% This would be of great benefit for further purposes since a calibration of the kinetic temperature from equilibrium measurements is a less data-demanding technique to characterize the thermodynamic processes.
%%% AO No estoy seguro de la segunda mitad de este párrafo. Podría ser como una conclusión de la sección. 

\begin{figure}
 \includegraphics[width=7cm]{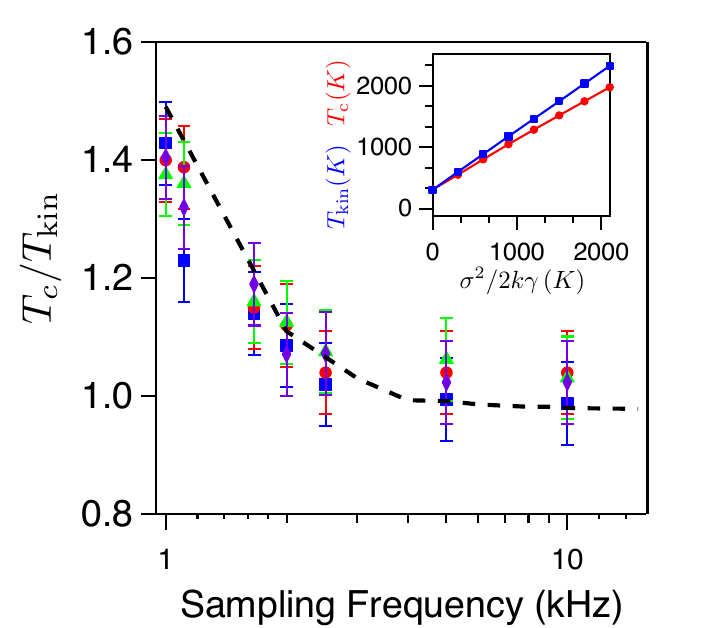}
 \caption{Values of the quotient $T_c/T_{\rm kin}$ in the dragging experiment as functions of the sampling frequency for different values of the external field, corresponding to the kinetic temperatures: $T_{\rm kin} = 525\,\rm K$ (blue squares), $T_{\rm kin} = 775 \,\rm K$ (red circles), $T_{\rm kin} = 1010\,\rm K$ (green triangles) and $T_{\rm kin}=1520\,\rm K$ (magenta diamonds). We also show the value of  $T_c/T_{\rm kin}$ as a function of the sampling frequency obtained from numerical simulations of the overdamped Langevin equation for an external noise with flat spectrum up to $f_{\rm co} = 3\,\rm kHz$ and intensity $\sigma^2/2k\gamma = 500\,\rm K$ (black  dashed curve). Inset: $T_{\rm kin}$ (blue squares) and  $T_{c}$ (red circles) as a function of noise intensity, $\sigma^2/2k\gamma$, for the experimental values of the experiment described in Ref.~\cite{martinez2013effective}. Solid lines are included to guide the eye.}
 \label{fig:Tc/Tkin}
\end{figure}

In order to clarify this issue, we now compare the values of $T_{\rm kin}$ and $T_c$ obtained for different values of the noise amplitude and different acquisition frequencies, ranging from $1\,\rm kHz$ to $10\,\rm kHz$. Figure~\ref{fig: figure4} shows that equilibrium and nonequilibrium effective temperatures do coincide within experimental errors when the sampling rate exceeds $f=2\,\rm kHz$. $T_{\rm kin}$ is measured from the variance of the position of the particle [Eq.~\eqref{eq: equilibrium}] from a time series of $20\,\rm s$ in which the trap is held fixed, yielding the very same value in the analyzed range of sampling frequency. When changing the position acquisition frequency, the value of $T_{\rm kin}$ does not change, whereas $T_c$ changes significantly up to a saturating value, reached when $f\simeq 2\,\rm kHz$. This deviation between equilibrium and nonequilibrium kinetic temperatures for certain values of the data acquisition rate could be a drawback for our setup to be applicable to design nonequilibrium thermodynamic processes at the mesoscale. 

We can get a deeper understanding of the mismatch between $T_{\rm kin}$ and $T_c$ by simulating the overdamped Langevin equation~\eqref{langevin} and taking into account the cutoff of the random force at $f_{\rm co} = 3\,\rm kHz$ recently observed in Ref.~\cite{roldan2014measuring}. Although the cutoff of the electric generator is in $10\,\rm kHz$, the force on the particle at frequencies above $3\,\rm kHz$ decays rapidly due to a relaxation of the polarization state of the particle and its electric double layer~\cite{rica2012electrokinetics,roldan2014measuring}. We investigate if the difference between $T_c$ and $T_{\rm kin}$ at low sampling frequencies can be assessed by our model. An analytical calculation of $T_c$ is cumbersome and can only be done for specific distributions of the random external noise, such as Gaussian white noise or Ornstein-Uhlenbeck noise~\cite{martinez2013effective,saha2009entropy}. We adopt a different but equivalent approach, performing numerical simulations of the overdamped Langevin equation~\eqref{langevin} using Euler numerical simulation scheme, with a simulation time step of $\Delta t=10^{-3} \,\rm ms$. The values of all the external parameters are set to those of the experiment. The spectrum of the external force is flat  up to a cutoff frequency of $f_{\rm co} = 3\,\rm kHz$ and its amplitude is arbitrarily set to a value $\sigma$ such that $\sigma^2/2k\gamma=500\,\rm K$. Such random force was attained by generating a Gaussian white noise signal and applying a filter with a cutoff frequency $f_{\rm co} = 3\,\rm kHz$ and followed by an inverse Fourier transform.

Figure~\ref{fig:Tc/Tkin} shows the values of the quotient $T_c/T_{\rm kin}$ as a function of the sampling frequency plotted for different values of the external field (different symbols in the Figure) corresponding of those described in the caption of Fig.~\ref{fig: figure4}. The solid black line in Fig.~\ref{fig:Tc/Tkin} shows that the value of $T_c/T_{\rm kin}$ as a function of the sampling frequency, as obtained from the numerical simulations, is in well agreement with the experimental measurements. The results in Fig.~\ref{fig:Tc/Tkin} indicate that sampling at frequencies above the noise cutoff frequency does not yield any difference in the nonequilibrium measurements. When sampling close to the corner frequency of the trap, $f_c=\frac{\kappa}{2\pi\gamma} = 340\,\rm Hz$ in this case, equilibrium and nonequilibrium kinetic temperatures do not coincide, and $T_c$ is above its equilibrium counterpart, $T_c>T_{\rm kin}$.

Interestingly, the opposite result $(T_c< T_{\rm kin})$ was reported in Ref.~\cite{martinez2013effective} for a similar dragging trap experiment. From the experimental point of view, we may note that in the present work, the noise cutoff frequency given by the amplifier is one order of magnitud larger than the one in Ref.~\cite{martinez2013effective}, and therefore the drawbacks of a colored spectrum of the noise are reduced~\cite{roldan2014measuring}. In the inset in Fig.~\ref{fig:Tc/Tkin}, we show that our model predicts this different behavior when using the experimental data in the experiment in Ref.~\cite{martinez2013effective} ($f_{\rm co}=1\,\rm kHz$, $\kappa=6\,$pN/$\mu$m, $\tau_1=\tau_2=6.3\,\rm ms$ and  $L=122\,\rm nm$ for instance). Therefore, the relation between $T_c$ and $T_{\rm kin}$ is complex and very sensitive to the values of the experimental parameters. 

From this discussion, we can conclude that a sampling frequency $f=2\,\rm kHz$ is optimal for the experiment we describe next, since it is below any relaxation of the external force ($\sim 3\,\rm kHz$), above the corner frequency ($\sim 300\,\rm Hz$), and does not unnecessarily store redundant data.

\begin{figure}[h!]
 \includegraphics[width=7cm]{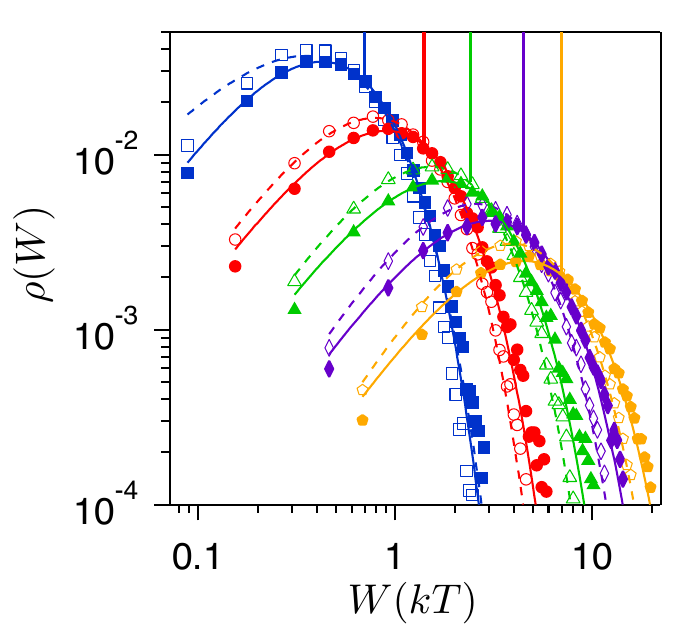}
 \caption{Work distributions in the isothermal compression [$\rho (W)$, filled symbols] and isothermal expansion [$\tilde\rho(-W)$, open symbols] for different values of the noise intensities corresponding to the following nonequilibrium effective temperatures: Without external field, $T_c=300\,\rm K$ (blue squares), $T_c=610\,\rm K$ (red circles), $T_c=885\,\rm K$ (green triangles), $T_c=1920\,\rm K$ (magenta diamonds) and $T_c=2950\,\rm K$ (orange pentagons). Solid and dashed curves are fits to Eq.~\eqref{eq:gammafit}. Vertical lines of the corresponding color show the expected value for the free energy change at the given temperatures. Data acquisition rate to calculate the work: $f=2\,\rm kHz$. }
 \label{fig: figure6}
\end{figure}

\begin{figure}[h!]
 \includegraphics[width=7cm]{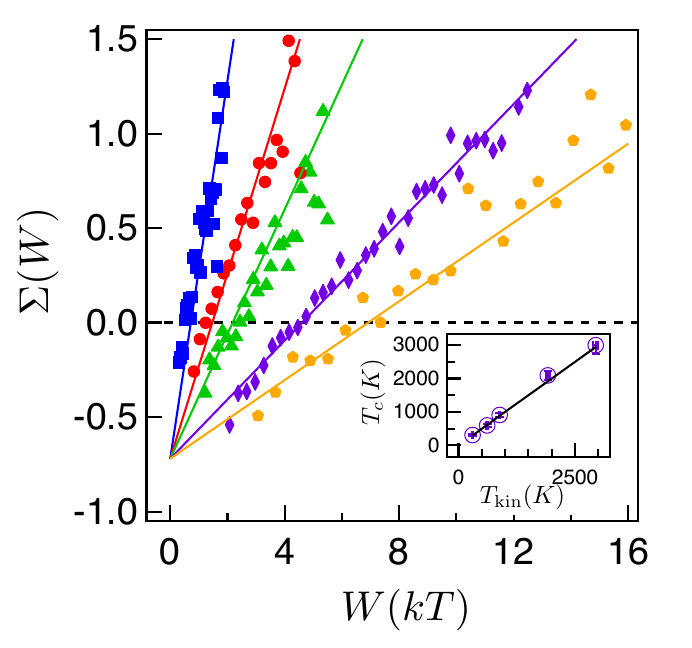}
 \caption{Experimental values (markers) and theoretical values (solid lines) of the work asymmetry function obtained from the work distributions in Fig.~\ref{fig: figure6}. %Red circles correspond to room temperature $T_c=300\,\rm K$, Orange stars to $T_c=610\,\rm K$, Green squares to $T_c=885\,\rm K$, cyan diamonds to $T_c=1920\,\rm K$  and blue triangles to $T_c=2950\,\rm K$. 
 Theoretical curves are computed using the values obtained for $T_{\rm kin}$. Inset: $T_c$ as a function of $T_{\rm kin}$ (open magenta circles, error bars are smaller than the symbol size). The solid line has slope $1$.}
 \label{fig: figure7}
\end{figure}

\subsection{Isothermal compression and expansion}

The distributions of the work (minus the work) in the forward (backward) process of increasing and decreasing the stiffness of the trap (see Fig.~\ref{fig:breathprotocol}) for different values of the external noise amplitude are shown in Fig.~\ref{fig: figure6}. We notice that the work fluctuations are non-Gaussian for both isothermal compression and expansion, as predicted theoretically in \cite{speck2011work}. The distributions can be fitted with a very good agreement to generalized Gamma distributions,
\begin{eqnarray}
\rho(W) &=& C_F \, W^{z_F} \, e^{-W/\alpha_F}, \label{eq:gammafit} \\
\tilde \rho(-W) &=& C_B \, (-W)^{z_B} \, e^{W/\alpha_B},  \label{eq:gammafitb}
\end{eqnarray}
where the fitting parameters $C_F$, $C_B$, $\alpha_F$ and $\alpha_B$ depend on the amplitude of the external noise, but not $z_F$ and $z_B$ (data not shown). The above result can be justified provided that the work along isothermal compression and expansion is equal to the sum of squared Gaussian variables [see Eq.~\eqref{eq:workbreath}] which is Gamma-distributed~\cite{sung2007theoretical,van1992stochastic}. Interestingly, the distributions~\eqref{eq:gammafit} and~\eqref{eq:gammafitb} are analogous to that of the work in the adiabatic compression or expansion of a dilute gas~\cite{crooks2007work}. 

The asymmetry between forward (compression) and backward (expansion) work distributions is an indicator of the irreversibility or the nonequilibrium nature of the process. In Fig.~\ref{fig: figure6} we show that the forward and backward work histograms cross at the value of the effective free energy change $\Delta F = kT_{\rm kin}\ln\sqrt{\kappa_{\rm fin}/\kappa_{\rm ini}}$ in all cases, with $T_{\rm kin}$ equal to the equilibrium kinetic temperature, measured in an independent equilibrium experiment. The difference between $\rho(W)$ and $\tilde \rho(-W)$ is quantified with the work asymmetry function, Eq.~\eqref{eq:asymmetryfunction}, whose values for different noise intensities are shown in Fig.~\ref{fig: figure7}. The work asymmetry function depends linearly on the work, with its slope equal to $1/kT_{\rm kin}$, or equivalently $T_c=T_{\rm kin}$. The inset in Fig.~\ref{fig: figure7} shows that this equality holds throughout the range of temperatures we explored. This result implies that our setup is suitable to implement nonequilibrium isothermal compression or expansions in the mesoscale, with the externally controlled temperature verifying all the requirements of an actual one.

%Work distributions of the breathing parabola are shown in Fig. \ref{fig: figure6}. These are exponentially distributed and keep only positive values since $\langle x^2 \rangle>0$ and the control parameter $\text{d}\kappa$ maintains its sign during each process. As in our first case of study, when noise amplitude increases, the width of the work distribution also increases. Moreover, it also induces a shift of the both work distributions (forward and backwards) to higher values of $\Delta F$ in agreement with eq.(\ref{eq:CFTcolor}). From CFT we can define an effective temperature for the free energy difference:
%
%\begin{equation}
%T_{\Delta F}=\frac{1}{k} \frac{W^{*}}{\beta \Delta F}
%\end{equation}
%
%that in our experiment satisfies $T_{\rm kin}\simeq T_{C} \simeq T_{\Delta F}$ showing consistency on the temperatures that the system feels at equilibrium for $\rho(x)$ and out of equilibrium for $W$ and $\Delta F$. 

%
%This is better shown in Figure \ref{fig: figure7}, where we plot the experimental values for $\Sigma(W)$ and we see that both $T_{C}$ (1/slope ) and  $T_{\Delta F}$ ($x$ axis and CFT curve crossing point) match well the predicted values for the CFT  at $T_{\rm kin}$. Figure \ref{fig: figure7} inset compares again the values $T_{\Delta F}$ vs. $T_{\rm kin}$ showing how random noise can mimic a thermal baths at temperatures up to several thousands of kelvins.

\section{Conclusions}
\label{sec:conclusions}

In this paper we have studied the dynamics of an optically-trapped microsphere immersed in water and subject to an external colored noise with flat spectrum up to a finite cutoff frequency. We have shown that, under these conditions, the fluctuations of the work in a nonequilibrium process define a temperature that coincides with the kinetic temperature of a particle in a thermal bath as obtained from equilibrium measurements. 
%%% AO No entendí esto.
This fact has been tested experimentally in two different nonequilibrium processes: first, dragging the trap at constant speed, and second, changing the trap stiffness linearly with time. In the second case we have found that the work fluctuations are non-Gaussian and fit well to a generalized gamma function.

The agreement between the temperature obtained from work fluctuations under a nonequilibrium driving ant the kinetic temperature obtained in equilibrium is only found when the work is calculated using a sampling rate significantly greater than the corner frequency of the trap. In a recent result~\cite{roldan2014measuring}, it has been shown that average kinetic energy changes in the mesoscale can be extrapolated from position samplings below the electrophoretic cutoff frequency. Such finding, together with the results of this article, suggest that an optimal sampling frequency to ascertain the complete energetics of a microsystem would be between the corner frequency of the trap and the electrophoretic cutoff frequency, $f=2\,\rm kHz$ being an optimal choice for the experimental conditions of the present work.

The main application of the experimental setup we introduced will be the construction of thermodynamic heat engines at the mesoscale where the temperature of the system can be arbitrarily switched. This opens the possibility for the design of nonequilibrium adiabatic processes or even the design of micro or nano Carnot engines, following the theoretical proposals in~\cite{hondou2000inattainability,esposito2010efficiency,schmiedl2008efficiency,rana2014single,verley2014unlikely}.

\section{Acknowledgments}
We acknowledge fruitful discussions with J. M. R. Parrondo, Luis Dinis and Dmitry Petrov. All the authors acknowledge financial support from the Fundaci\'o Privada Cellex Barcelona, Generalitat de Catalunya grant 2009-SGR-159, and from the MICINN (grant FIS2011-24409).
ER acknowledges financial support from the Spanish Government (ENFASIS) and Max Planck Institute for the Physics of Complex Systems. 
AO acknowledges support from the Concejo Nacional de Ciencia y Tecnolog\'ia and from Tecnol\'ogico de Monterrey (grant CAT-141).


\begin{thebibliography}{43}
\expandafter\ifx\csname natexlab\endcsname\relax\def\natexlab#1{#1}\fi
\expandafter\ifx\csname bibnamefont\endcsname\relax
  \def\bibnamefont#1{#1}\fi
\expandafter\ifx\csname bibfnamefont\endcsname\relax
  \def\bibfnamefont#1{#1}\fi
\expandafter\ifx\csname citenamefont\endcsname\relax
  \def\citenamefont#1{#1}\fi
\expandafter\ifx\csname url\endcsname\relax
  \def\url#1{\texttt{#1}}\fi
\expandafter\ifx\csname urlprefix\endcsname\relax\def\urlprefix{URL }\fi
\providecommand{\bibinfo}[2]{#2}
\providecommand{\eprint}[2][]{\url{#2}}

\bibitem[{\citenamefont{Seifert}(2012)}]{seifert2012stochastic}
\bibinfo{author}{\bibfnamefont{U.}~\bibnamefont{Seifert}},
  \bibinfo{journal}{Rep. Prog. Phys.} \textbf{\bibinfo{volume}{75}},
  \bibinfo{pages}{126001} (\bibinfo{year}{2012}).

\bibitem[{\citenamefont{Volpe et~al.}(2008)\citenamefont{Volpe, Perrone, Rubi,
  and Petrov}}]{volpe2008stochastic}
\bibinfo{author}{\bibfnamefont{G.}~\bibnamefont{Volpe}},
  \bibinfo{author}{\bibfnamefont{S.}~\bibnamefont{Perrone}},
  \bibinfo{author}{\bibfnamefont{J.~M.} \bibnamefont{Rubi}}, \bibnamefont{and}
  \bibinfo{author}{\bibfnamefont{D.}~\bibnamefont{Petrov}},
  \bibinfo{journal}{Phys. Rev. E} \textbf{\bibinfo{volume}{77}},
  \bibinfo{pages}{051107} (\bibinfo{year}{2008}).

\bibitem[{\citenamefont{Wang et~al.}(2002)\citenamefont{Wang, Sevick, Mittag,
  Searles, and Evans}}]{wang2002experimental}
\bibinfo{author}{\bibfnamefont{G.}~\bibnamefont{Wang}},
  \bibinfo{author}{\bibfnamefont{E.~M.} \bibnamefont{Sevick}},
  \bibinfo{author}{\bibfnamefont{E.}~\bibnamefont{Mittag}},
  \bibinfo{author}{\bibfnamefont{D.~J.} \bibnamefont{Searles}},
  \bibnamefont{and} \bibinfo{author}{\bibfnamefont{D.~J.} \bibnamefont{Evans}},
  \bibinfo{journal}{Phys. Rev. Lett.} \textbf{\bibinfo{volume}{89}},
  \bibinfo{pages}{050601} (\bibinfo{year}{2002}).

\bibitem[{\citenamefont{Blickle and Bechinger}(2012)}]{blickle2012realization}
\bibinfo{author}{\bibfnamefont{V.}~\bibnamefont{Blickle}} \bibnamefont{and}
  \bibinfo{author}{\bibfnamefont{C.}~\bibnamefont{Bechinger}},
  \bibinfo{journal}{Nature Phys.} \textbf{\bibinfo{volume}{8}},
  \bibinfo{pages}{143} (\bibinfo{year}{2012}).

\bibitem[{\citenamefont{Yasuda et~al.}(1998)\citenamefont{Yasuda, Noji,
  Kinosita~Jr, and Yoshida}}]{yasuda1998f}
\bibinfo{author}{\bibfnamefont{R.}~\bibnamefont{Yasuda}},
  \bibinfo{author}{\bibfnamefont{H.}~\bibnamefont{Noji}},
  \bibinfo{author}{\bibfnamefont{K.}~\bibnamefont{Kinosita~Jr}},
  \bibnamefont{and} \bibinfo{author}{\bibfnamefont{M.}~\bibnamefont{Yoshida}},
  \bibinfo{journal}{Cell} \textbf{\bibinfo{volume}{93}}, \bibinfo{pages}{1117}
  (\bibinfo{year}{1998}).

\bibitem[{\citenamefont{Mao et~al.}(2005)\citenamefont{Mao, Ricardo
  Arias-Gonzalez, Smith, Tinoco~Jr, and Bustamante}}]{mao2005temperature}
\bibinfo{author}{\bibfnamefont{H.}~\bibnamefont{Mao}},
  \bibinfo{author}{\bibfnamefont{J.}~\bibnamefont{Ricardo Arias-Gonzalez}},
  \bibinfo{author}{\bibfnamefont{S.~B.} \bibnamefont{Smith}},
  \bibinfo{author}{\bibfnamefont{I.}~\bibnamefont{Tinoco~Jr}},
  \bibnamefont{and}
  \bibinfo{author}{\bibfnamefont{C.}~\bibnamefont{Bustamante}},
  \bibinfo{journal}{Biophys. J.} \textbf{\bibinfo{volume}{89}},
  \bibinfo{pages}{1308} (\bibinfo{year}{2005}).

\bibitem[{\citenamefont{Millen et~al.}(2014)\citenamefont{Millen, Deesuwan,
  Barker, and Anders}}]{millen2014nanoscale}
\bibinfo{author}{\bibfnamefont{J.}~\bibnamefont{Millen}},
  \bibinfo{author}{\bibfnamefont{T.}~\bibnamefont{Deesuwan}},
  \bibinfo{author}{\bibfnamefont{P.}~\bibnamefont{Barker}}, \bibnamefont{and}
  \bibinfo{author}{\bibfnamefont{J.}~\bibnamefont{Anders}},
  \bibinfo{journal}{Nature Nanotech.} \textbf{\bibinfo{volume}{9}},
  \bibinfo{pages}{425} (\bibinfo{year}{2014}).

\bibitem[{\citenamefont{Sekimoto}(2010)}]{sekimoto-bible}
\bibinfo{author}{\bibfnamefont{K.}~\bibnamefont{Sekimoto}}, in
  \emph{\bibinfo{booktitle}{Lecture Notes in Physics, Berlin Springer Verlag}}
  (\bibinfo{year}{2010}), vol. \bibinfo{volume}{799}.

\bibitem[{\citenamefont{Gomez-Solano et~al.}(2010)\citenamefont{Gomez-Solano,
  Bellon, Petrosyan, and Ciliberto}}]{gomez2010steady}
\bibinfo{author}{\bibfnamefont{J.~R.} \bibnamefont{Gomez-Solano}},
  \bibinfo{author}{\bibfnamefont{L.}~\bibnamefont{Bellon}},
  \bibinfo{author}{\bibfnamefont{A.}~\bibnamefont{Petrosyan}},
  \bibnamefont{and}
  \bibinfo{author}{\bibfnamefont{S.}~\bibnamefont{Ciliberto}},
  \bibinfo{journal}{EPL--Europhys. Lett.} \textbf{\bibinfo{volume}{89}},
  \bibinfo{pages}{60003} (\bibinfo{year}{2010}).

\bibitem[{\citenamefont{Mart{\'\i}nez et~al.}(2013)\citenamefont{Mart{\'\i}nez,
  Rold{\'a}n, Parrondo, and Petrov}}]{martinez2013effective}
\bibinfo{author}{\bibfnamefont{I.~A.} \bibnamefont{Mart{\'\i}nez}},
  \bibinfo{author}{\bibfnamefont{{\'E}.}~\bibnamefont{Rold{\'a}n}},
  \bibinfo{author}{\bibfnamefont{J.~M.} \bibnamefont{Parrondo}},
  \bibnamefont{and} \bibinfo{author}{\bibfnamefont{D.}~\bibnamefont{Petrov}},
  \bibinfo{journal}{Phys. Rev. E} \textbf{\bibinfo{volume}{87}},
  \bibinfo{pages}{032159} (\bibinfo{year}{2013}).

\bibitem[{\citenamefont{Crooks}(1999)}]{crooks1999entropy}
\bibinfo{author}{\bibfnamefont{G.~E.} \bibnamefont{Crooks}},
  \bibinfo{journal}{Phys. Rev. E} \textbf{\bibinfo{volume}{60}},
  \bibinfo{pages}{2721} (\bibinfo{year}{1999}).

\bibitem[{\citenamefont{Ashkin}(1970)}]{ashkin1970acceleration}
\bibinfo{author}{\bibfnamefont{A.}~\bibnamefont{Ashkin}},
  \bibinfo{journal}{Phys. Rev. Lett.} \textbf{\bibinfo{volume}{24}},
  \bibinfo{pages}{156} (\bibinfo{year}{1970}).

\bibitem[{\citenamefont{Evers et~al.}(2013)\citenamefont{Evers, Zunke, Hanes,
  Bewerunge, Ladadwa, Heuer, and Egelhaaf}}]{evers2013particle}
\bibinfo{author}{\bibfnamefont{F.}~\bibnamefont{Evers}},
  \bibinfo{author}{\bibfnamefont{C.}~\bibnamefont{Zunke}},
  \bibinfo{author}{\bibfnamefont{R.~D.~L.} \bibnamefont{Hanes}},
  \bibinfo{author}{\bibfnamefont{J.}~\bibnamefont{Bewerunge}},
  \bibinfo{author}{\bibfnamefont{I.}~\bibnamefont{Ladadwa}},
  \bibinfo{author}{\bibfnamefont{A.}~\bibnamefont{Heuer}}, \bibnamefont{and}
  \bibinfo{author}{\bibfnamefont{S.~U.} \bibnamefont{Egelhaaf}},
  \bibinfo{journal}{Phys. Rev. E} \textbf{\bibinfo{volume}{88}},
  \bibinfo{pages}{022125} (\bibinfo{year}{2013}).

\bibitem[{\citenamefont{Simon et~al.}(2013)\citenamefont{Simon, Sancho, and
  Lindenberg}}]{simon2013transport}
\bibinfo{author}{\bibfnamefont{M.~S.~n.} \bibnamefont{Simon}},
  \bibinfo{author}{\bibfnamefont{J.~M.} \bibnamefont{Sancho}},
  \bibnamefont{and}
  \bibinfo{author}{\bibfnamefont{K.}~\bibnamefont{Lindenberg}},
  \bibinfo{journal}{Phys. Rev. E} \textbf{\bibinfo{volume}{88}},
  \bibinfo{pages}{062105} (\bibinfo{year}{2013}).

\bibitem[{\citenamefont{Gieseler et~al.}(2014)\citenamefont{Gieseler, Quidant,
  Dellago, and Novotny}}]{gieseler2014dynamic}
\bibinfo{author}{\bibfnamefont{J.}~\bibnamefont{Gieseler}},
  \bibinfo{author}{\bibfnamefont{R.}~\bibnamefont{Quidant}},
  \bibinfo{author}{\bibfnamefont{C.}~\bibnamefont{Dellago}}, \bibnamefont{and}
  \bibinfo{author}{\bibfnamefont{L.}~\bibnamefont{Novotny}},
  \bibinfo{journal}{Nature Nanotech.} \textbf{\bibinfo{volume}{9}},
  \bibinfo{pages}{358} (\bibinfo{year}{2014}).

\bibitem[{\citenamefont{Volpe et~al.}(2014)\citenamefont{Volpe, Kurz,
  Callegari, Volpe, and Gigan}}]{volpe2014speckle}
\bibinfo{author}{\bibfnamefont{G.}~\bibnamefont{Volpe}},
  \bibinfo{author}{\bibfnamefont{L.}~\bibnamefont{Kurz}},
  \bibinfo{author}{\bibfnamefont{A.}~\bibnamefont{Callegari}},
  \bibinfo{author}{\bibfnamefont{G.}~\bibnamefont{Volpe}}, \bibnamefont{and}
  \bibinfo{author}{\bibfnamefont{S.}~\bibnamefont{Gigan}}
  (\bibinfo{year}{2014}), \bibinfo{note}{arXiv prepint - arXiv 1403.0364}.

\bibitem[{\citenamefont{Rold\'an}(2014)}]{roldan2014irreversibility}
\bibinfo{author}{\bibfnamefont{E.}~\bibnamefont{Rold\'an}},
  \emph{\bibinfo{title}{Irreversibility and dissipation in microscopic
  systems}} (\bibinfo{publisher}{Springer Theses}, \bibinfo{address}{Berlin},
  \bibinfo{year}{2014}), ISBN \bibinfo{isbn}{978-3-319-07079-7}.

\bibitem[{\citenamefont{Raj et~al.}(2012)\citenamefont{Raj, Marro, Wojdyla, and
  Petrov}}]{raj2012mechanochemistry}
\bibinfo{author}{\bibfnamefont{S.}~\bibnamefont{Raj}},
  \bibinfo{author}{\bibfnamefont{M.}~\bibnamefont{Marro}},
  \bibinfo{author}{\bibfnamefont{M.}~\bibnamefont{Wojdyla}}, \bibnamefont{and}
  \bibinfo{author}{\bibfnamefont{D.}~\bibnamefont{Petrov}},
  \bibinfo{journal}{Biomed. Opt. Expr.} \textbf{\bibinfo{volume}{3}},
  \bibinfo{pages}{753} (\bibinfo{year}{2012}), \bibinfo{note}{cited By (since
  1996)6}.

\bibitem[{\citenamefont{Rao et~al.}(2013)\citenamefont{Rao, Raj, Cossins,
  Marro, Guallar, and Petrov}}]{rao2013direct}
\bibinfo{author}{\bibfnamefont{S.}~\bibnamefont{Rao}},
  \bibinfo{author}{\bibfnamefont{S.}~\bibnamefont{Raj}},
  \bibinfo{author}{\bibfnamefont{B.}~\bibnamefont{Cossins}},
  \bibinfo{author}{\bibfnamefont{M.}~\bibnamefont{Marro}},
  \bibinfo{author}{\bibfnamefont{V.}~\bibnamefont{Guallar}}, \bibnamefont{and}
  \bibinfo{author}{\bibfnamefont{D.}~\bibnamefont{Petrov}},
  \bibinfo{journal}{Biophys. J.} \textbf{\bibinfo{volume}{104}},
  \bibinfo{pages}{156} (\bibinfo{year}{2013}).

\bibitem[{\citenamefont{Gross et~al.}(2011)\citenamefont{Gross, Laurens,
  Oddershede, Bockelmann, Peterman, and Wuite}}]{gross2011quantifying}
\bibinfo{author}{\bibfnamefont{P.}~\bibnamefont{Gross}},
  \bibinfo{author}{\bibfnamefont{N.}~\bibnamefont{Laurens}},
  \bibinfo{author}{\bibfnamefont{L.~B.} \bibnamefont{Oddershede}},
  \bibinfo{author}{\bibfnamefont{U.}~\bibnamefont{Bockelmann}},
  \bibinfo{author}{\bibfnamefont{E.~J.} \bibnamefont{Peterman}},
  \bibnamefont{and} \bibinfo{author}{\bibfnamefont{G.~J.} \bibnamefont{Wuite}},
  \bibinfo{journal}{Nature Phys.} \textbf{\bibinfo{volume}{7}},
  \bibinfo{pages}{731} (\bibinfo{year}{2011}).

\bibitem[{\citenamefont{Lu et~al.}(2012)\citenamefont{Lu, Terray, Collins, and
  Hart}}]{lu2012single}
\bibinfo{author}{\bibfnamefont{Q.}~\bibnamefont{Lu}},
  \bibinfo{author}{\bibfnamefont{A.}~\bibnamefont{Terray}},
  \bibinfo{author}{\bibfnamefont{G.~E.} \bibnamefont{Collins}},
  \bibnamefont{and} \bibinfo{author}{\bibfnamefont{S.~J.} \bibnamefont{Hart}},
  \bibinfo{journal}{Lab Chip} \textbf{\bibinfo{volume}{12}},
  \bibinfo{pages}{1128} (\bibinfo{year}{2012}).

\bibitem[{\citenamefont{Semenov et~al.}(2013)\citenamefont{Semenov, Raafatnia,
  Sega, Lobaskin, Holm, and Kremer}}]{semenov2013electrophoretic}
\bibinfo{author}{\bibfnamefont{I.}~\bibnamefont{Semenov}},
  \bibinfo{author}{\bibfnamefont{S.}~\bibnamefont{Raafatnia}},
  \bibinfo{author}{\bibfnamefont{M.}~\bibnamefont{Sega}},
  \bibinfo{author}{\bibfnamefont{V.}~\bibnamefont{Lobaskin}},
  \bibinfo{author}{\bibfnamefont{C.}~\bibnamefont{Holm}}, \bibnamefont{and}
  \bibinfo{author}{\bibfnamefont{F.}~\bibnamefont{Kremer}},
  \bibinfo{journal}{Phys. Rev. E} \textbf{\bibinfo{volume}{87}},
  \bibinfo{pages}{022302} (\bibinfo{year}{2013}).

\bibitem[{\citenamefont{Strubbe et~al.}(2013)\citenamefont{Strubbe, Beunis,
  Brans, Karvar, Woestenborghs, and Neyts}}]{strubbe2013electrophoretic}
\bibinfo{author}{\bibfnamefont{F.}~\bibnamefont{Strubbe}},
  \bibinfo{author}{\bibfnamefont{F.}~\bibnamefont{Beunis}},
  \bibinfo{author}{\bibfnamefont{T.}~\bibnamefont{Brans}},
  \bibinfo{author}{\bibfnamefont{M.}~\bibnamefont{Karvar}},
  \bibinfo{author}{\bibfnamefont{W.}~\bibnamefont{Woestenborghs}},
  \bibnamefont{and} \bibinfo{author}{\bibfnamefont{K.}~\bibnamefont{Neyts}},
  \bibinfo{journal}{Phys. Rev. X} \textbf{\bibinfo{volume}{3}},
  \bibinfo{pages}{021001} (\bibinfo{year}{2013}).

\bibitem[{\citenamefont{Jonás and Zemánek}(2008)}]{jonas2008light}
\bibinfo{author}{\bibfnamefont{A.}~\bibnamefont{Jonás}} \bibnamefont{and}
  \bibinfo{author}{\bibfnamefont{P.}~\bibnamefont{Zemánek}},
  \bibinfo{journal}{Electrophoresis} \textbf{\bibinfo{volume}{29}},
  \bibinfo{pages}{4813} (\bibinfo{year}{2008}), ISSN \bibinfo{issn}{1522-2683}.

\bibitem[{\citenamefont{Rold\'an et~al.}(2014)\citenamefont{Rold\'an,
  Mart\'inez, Parrondo, and Petrov}}]{Roldan2014Universal}
\bibinfo{author}{\bibfnamefont{E.}~\bibnamefont{Rold\'an}},
  \bibinfo{author}{\bibfnamefont{I.~A.} \bibnamefont{Mart\'inez}},
  \bibinfo{author}{\bibfnamefont{J.~M.~R.} \bibnamefont{Parrondo}},
  \bibnamefont{and} \bibinfo{author}{\bibfnamefont{D.}~\bibnamefont{Petrov}},
  \bibinfo{journal}{Nature Phys.} \textbf{\bibinfo{volume}{10}},
  \bibinfo{pages}{457} (\bibinfo{year}{2014}).

\bibitem[{\citenamefont{Rold{\'a}n et~al.}(2014)\citenamefont{Rold{\'a}n,
  Mart{\'\i}nez, Dinis, and Rica}}]{roldan2014measuring}
\bibinfo{author}{\bibfnamefont{{\'E}.}~\bibnamefont{Rold{\'a}n}},
  \bibinfo{author}{\bibfnamefont{I.~A.} \bibnamefont{Mart{\'\i}nez}},
  \bibinfo{author}{\bibfnamefont{L.}~\bibnamefont{Dinis}}, \bibnamefont{and}
  \bibinfo{author}{\bibfnamefont{R.~A.} \bibnamefont{Rica}},
  \bibinfo{journal}{Appl. Phys. Lett.} \textbf{\bibinfo{volume}{104}},
  \bibinfo{pages}{234103} (\bibinfo{year}{2014}).

\bibitem[{\citenamefont{Mart\'inez et~al.}(2014)\citenamefont{Mart\'inez,
  Rold\'an, Dinis, Rica, and Petrov}}]{martinez2014realization}
\bibinfo{author}{\bibfnamefont{I.~A.} \bibnamefont{Mart\'inez}},
  \bibinfo{author}{\bibfnamefont{E.}~\bibnamefont{Rold\'an}},
  \bibinfo{author}{\bibfnamefont{L.}~\bibnamefont{Dinis}},
  \bibinfo{author}{\bibfnamefont{R.~A.} \bibnamefont{Rica}}, \bibnamefont{and}
  \bibinfo{author}{\bibfnamefont{D.}~\bibnamefont{Petrov}}
  (\bibinfo{year}{2014}), \bibinfo{note}{under review}.

\bibitem[{\citenamefont{Langevin}(1908)}]{Langevin1908}
\bibinfo{author}{\bibfnamefont{P.}~\bibnamefont{Langevin}},
  \bibinfo{journal}{CR Acad. Sci. Paris} \textbf{\bibinfo{volume}{146}}
  (\bibinfo{year}{1908}).

\bibitem[{\citenamefont{Garnier and
  Ciliberto}(2005)}]{garnier2005nonequilibrium}
\bibinfo{author}{\bibfnamefont{N.}~\bibnamefont{Garnier}} \bibnamefont{and}
  \bibinfo{author}{\bibfnamefont{S.}~\bibnamefont{Ciliberto}},
  \bibinfo{journal}{Phys. Rev. E} \textbf{\bibinfo{volume}{71}},
  \bibinfo{pages}{060101} (\bibinfo{year}{2005}).

\bibitem[{\citenamefont{Collin et~al.}(2005)\citenamefont{Collin, Ritort,
  Jarzynski, Smith, Tinoco, and Bustamante}}]{collin2005verification}
\bibinfo{author}{\bibfnamefont{D.}~\bibnamefont{Collin}},
  \bibinfo{author}{\bibfnamefont{F.}~\bibnamefont{Ritort}},
  \bibinfo{author}{\bibfnamefont{C.}~\bibnamefont{Jarzynski}},
  \bibinfo{author}{\bibfnamefont{S.}~\bibnamefont{Smith}},
  \bibinfo{author}{\bibfnamefont{I.}~\bibnamefont{Tinoco}}, \bibnamefont{and}
  \bibinfo{author}{\bibfnamefont{C.}~\bibnamefont{Bustamante}},
  \bibinfo{journal}{Nature} \textbf{\bibinfo{volume}{437}},
  \bibinfo{pages}{231} (\bibinfo{year}{2005}).

\bibitem[{\citenamefont{Tonin et~al.}(2010)\citenamefont{Tonin, B{\'a}lint,
  Mestres, Mart{\`\i}nez, and Petrov}}]{tonin2010electrophoretic}
\bibinfo{author}{\bibfnamefont{M.}~\bibnamefont{Tonin}},
  \bibinfo{author}{\bibfnamefont{S.}~\bibnamefont{B{\'a}lint}},
  \bibinfo{author}{\bibfnamefont{P.}~\bibnamefont{Mestres}},
  \bibinfo{author}{\bibfnamefont{I.~A.} \bibnamefont{Mart{\`\i}nez}},
  \bibnamefont{and} \bibinfo{author}{\bibfnamefont{D.}~\bibnamefont{Petrov}},
  \bibinfo{journal}{Applied Physics Letters} \textbf{\bibinfo{volume}{97}},
  \bibinfo{pages}{203704} (\bibinfo{year}{2010}).

\bibitem[{\citenamefont{Berg-S{\o}rensen and Flyvbjerg}(2004)}]{berg2004power}
\bibinfo{author}{\bibfnamefont{K.}~\bibnamefont{Berg-S{\o}rensen}}
  \bibnamefont{and}
  \bibinfo{author}{\bibfnamefont{H.}~\bibnamefont{Flyvbjerg}},
  \bibinfo{journal}{Rev. Sci. Instr.} \textbf{\bibinfo{volume}{75}},
  \bibinfo{pages}{594} (\bibinfo{year}{2004}).

\bibitem[{\citenamefont{Saha et~al.}(2009)\citenamefont{Saha, Lahiri, and
  Jayannavar}}]{saha2009entropy}
\bibinfo{author}{\bibfnamefont{A.}~\bibnamefont{Saha}},
  \bibinfo{author}{\bibfnamefont{S.}~\bibnamefont{Lahiri}}, \bibnamefont{and}
  \bibinfo{author}{\bibfnamefont{A.}~\bibnamefont{Jayannavar}},
  \bibinfo{journal}{Phys. Rev. E} \textbf{\bibinfo{volume}{80}},
  \bibinfo{pages}{011117} (\bibinfo{year}{2009}).

\bibitem[{\citenamefont{Rica et~al.}(2012)\citenamefont{Rica, Jim{\'e}nez, and
  Delgado}}]{rica2012electrokinetics}
\bibinfo{author}{\bibfnamefont{R.~A.} \bibnamefont{Rica}},
  \bibinfo{author}{\bibfnamefont{M.~L.} \bibnamefont{Jim{\'e}nez}},
  \bibnamefont{and} \bibinfo{author}{\bibfnamefont{{\'A}.~V.}
  \bibnamefont{Delgado}}, \bibinfo{journal}{Soft Matt.}
  \textbf{\bibinfo{volume}{8}}, \bibinfo{pages}{3596} (\bibinfo{year}{2012}).

\bibitem[{\citenamefont{Speck}(2011)}]{speck2011work}
\bibinfo{author}{\bibfnamefont{T.}~\bibnamefont{Speck}}, \bibinfo{journal}{J.
  Phys. A} \textbf{\bibinfo{volume}{44}}, \bibinfo{pages}{305001}
  (\bibinfo{year}{2011}).

\bibitem[{\citenamefont{Sung}(2007)}]{sung2007theoretical}
\bibinfo{author}{\bibfnamefont{J.}~\bibnamefont{Sung}}, \bibinfo{journal}{Phys.
  Rev. E} \textbf{\bibinfo{volume}{76}}, \bibinfo{pages}{012101}
  (\bibinfo{year}{2007}).

\bibitem[{\citenamefont{Van~Kampen}(1992)}]{van1992stochastic}
\bibinfo{author}{\bibfnamefont{N.~G.} \bibnamefont{Van~Kampen}},
  \emph{\bibinfo{title}{Stochastic processes in physics and chemistry}},
  vol.~\bibinfo{volume}{1} (\bibinfo{publisher}{Elsevier},
  \bibinfo{year}{1992}).

\bibitem[{\citenamefont{Crooks and Jarzynski}(2007)}]{crooks2007work}
\bibinfo{author}{\bibfnamefont{G.~E.} \bibnamefont{Crooks}} \bibnamefont{and}
  \bibinfo{author}{\bibfnamefont{C.}~\bibnamefont{Jarzynski}},
  \bibinfo{journal}{Phys. Rev. E} \textbf{\bibinfo{volume}{75}},
  \bibinfo{pages}{021116} (\bibinfo{year}{2007}).

\bibitem[{\citenamefont{Hondou and Sekimoto}(2000)}]{hondou2000inattainability}
\bibinfo{author}{\bibfnamefont{T.}~\bibnamefont{Hondou}} \bibnamefont{and}
  \bibinfo{author}{\bibfnamefont{K.}~\bibnamefont{Sekimoto}},
  \bibinfo{journal}{Phys. Rev. E} \textbf{\bibinfo{volume}{62}},
  \bibinfo{pages}{6021} (\bibinfo{year}{2000}).

\bibitem[{\citenamefont{Esposito et~al.}(2010)\citenamefont{Esposito, Kawai,
  Lindenberg, and Van~den Broeck}}]{esposito2010efficiency}
\bibinfo{author}{\bibfnamefont{M.}~\bibnamefont{Esposito}},
  \bibinfo{author}{\bibfnamefont{R.}~\bibnamefont{Kawai}},
  \bibinfo{author}{\bibfnamefont{K.}~\bibnamefont{Lindenberg}},
  \bibnamefont{and} \bibinfo{author}{\bibfnamefont{C.}~\bibnamefont{Van~den
  Broeck}}, \bibinfo{journal}{Phys. Rev. Lett.} \textbf{\bibinfo{volume}{105}},
  \bibinfo{pages}{150603} (\bibinfo{year}{2010}).

\bibitem[{\citenamefont{Schmiedl and Seifert}(2008)}]{schmiedl2008efficiency}
\bibinfo{author}{\bibfnamefont{T.}~\bibnamefont{Schmiedl}} \bibnamefont{and}
  \bibinfo{author}{\bibfnamefont{U.}~\bibnamefont{Seifert}},
  \bibinfo{journal}{EPL -- Europhys. Lett.} \textbf{\bibinfo{volume}{81}},
  \bibinfo{pages}{20003} (\bibinfo{year}{2008}).

\bibitem[{\citenamefont{Rana et~al.}(2014)\citenamefont{Rana, Pal, Saha, and
  Jayannavar}}]{rana2014single}
\bibinfo{author}{\bibfnamefont{S.}~\bibnamefont{Rana}},
  \bibinfo{author}{\bibfnamefont{P.}~\bibnamefont{Pal}},
  \bibinfo{author}{\bibfnamefont{A.}~\bibnamefont{Saha}}, \bibnamefont{and}
  \bibinfo{author}{\bibfnamefont{A.}~\bibnamefont{Jayannavar}},
  \bibinfo{journal}{arXiv preprint arXiv:1404.7831}  (\bibinfo{year}{2014}).

\bibitem[{\citenamefont{Verley et~al.}(2014)\citenamefont{Verley, Willaert,
  Van~de Broeck, and Esposito}}]{verley2014unlikely}
\bibinfo{author}{\bibfnamefont{G.}~\bibnamefont{Verley}},
  \bibinfo{author}{\bibfnamefont{T.}~\bibnamefont{Willaert}},
  \bibinfo{author}{\bibfnamefont{C.}~\bibnamefont{Van~de Broeck}},
  \bibnamefont{and} \bibinfo{author}{\bibfnamefont{M.}~\bibnamefont{Esposito}},
  \bibinfo{journal}{arXiv preprint arXiv:1404.3095}  (\bibinfo{year}{2014}).

\end{thebibliography}
\end{document}